\documentclass[runningheads]{llncs}
\usepackage{graphicx}
\usepackage{cite}
\usepackage{amsmath,amssymb,amsfonts}
\usepackage{graphicx}
\usepackage[caption=false]{subfig}
\usepackage{textcomp}
\usepackage{xcolor}
\usepackage{gensymb} % used for writing \degree
\usepackage{algorithm}

\usepackage{algorithmic}
\usepackage[multiple]{footmisc}

\usepackage[shortlabels]{enumitem}
\usepackage{comment} % used for comment
\usepackage{float} % used for subfloat
\usepackage{longtable} % used for longtable
\usepackage{enumitem} % used for removing indentation for bullet points

\begin{document}
\title{Dynamic Conflict Resolution of IoT Services in Smart Homes} %\vspace{-3em}}
%\thanks{Supported by organization x.}
%
\titlerunning{Dynamic Conflict Resolution of IoT Services in Smart Homes}

\author{Dipankar Chaki\inst{} \and
Athman Bouguettaya\inst{}}
\authorrunning{D. Chaki and A. Bouguettaya}
\institute{School of Computer Science, University of Sydney, Australia \\
%Springer Heidelberg, Tiergartenstr. 17, 69121 Heidelberg, Germany
\email{\{dipankar.chaki,athman.bouguettaya\}@sydney.edu.au}}

\maketitle              % typeset the header of the contribution
\begin{abstract}
We propose a novel conflict resolution framework for IoT services in multi-resident smart homes. The proposed framework employs a preference extraction model based on a temporal proximity strategy. We design a preference aggregation model using a matrix factorization-based approach (i.e., singular value decomposition). The concepts of current resident item matrix and ideal resident item matrix are introduced as key criteria to cater to the conflict resolution framework. Finally, a set of experiments on real-world datasets are conducted to show the effectiveness of the proposed approach.

%\vspace{-1em}

\end{abstract}

%\vspace{-3em}

\section{Introduction}
%\vspace{-3mm}
Internet of Things (IoT) is the umbrella term covering everyday objects (a.k.a. things) that are connected to the Internet. These are usually equipped with ubiquitous intelligence \cite{nauman2020multimedia}. 
%The rapid advancement of the underlying technologies such as wireless sensor networks, radio frequency identification tags, and barcodes provides augmented capabilities, including sensing, networking, and processing \cite{marwedel2021embedded}. 
IoT technologies are the key enablers of many cutting-edge applications such as smart cities, smart campuses, smart grids, and intelligent transport systems. A particular application domain of IoT is \emph{smart homes}. A smart home is defined as a home that is fitted with IoT devices. These IoT devices are attached to everyday ``things" to monitor usage patterns. The purpose of a smart home is to provide its residents with \textit{convenience} and \textit{efficiency} \cite{huang2018convenience}.

The concept of IoT is congruent with the \emph{service paradigm} \cite{bouguettaya2017service}. Each ``thing" has a set of \emph{functional} and \emph{non-functional} (a.k.a. quality of service) properties. In this regard, we leverage the service paradigm as a framework to define the functional and non-functional properties of smart home devices as \emph{IoT services} \cite{chaki2020fine}. For instance, a light bulb in a smart home is regarded as a light service. The functional property of the light service is to provide
illumination. Examples of non-functional properties include luminous intensity, color, connectivity.
%For instance, an Air-conditioning unit (AC) in a smart home is represented as an AC service. The functional property of the AC service is to control the temperature inside an ambient environment. Examples of non-functional properties include AC types (i.e., split-system, ducted, wall/window, portable), fan speed, fan direction, noise level, sleep mode, and dehumidifier mode.

%We identify two categories of smart homes: (i) \emph{Single resident} and (ii) \emph{Multi-resident}. \textit{IoT service conflicts} in a single resident home are fundamentally different from those in a multi-resident home. The focus of this paper is on IoT service conflicts that occur in multi-resident smart homes.
In a multi-occupant smart home, different residents may have different service requirements, leading to IoT service conflicts \cite{chaki2021adaptive}. For example, a resident may prefer the light to be ``on" while watching TV, and another resident may prefer the light to be ``off". Therefore, an \emph{IoT service conflict} occurs since the light service cannot satisfy multiple residents' requirements at the same time and location. In this context, \textit{detecting} and \textit{resolving} conflicts is paramount to provide residents with a higher level of convenience and satisfaction.

Residents usually communicate \emph{face-to-face} when co-located in a home. They can exchange their opinion and decide the appropriate state of shared services through this face-to-face communication. For example, family members may decide to watch a television channel by discussing with each other. Although this communication enables them to discuss their interest in television shows, it is cumbersome to find a show that would be agreeable to all in a world of thousands of available channels. This negotiation may lead to \emph{tension} and \emph{stress} \cite{shin2008mixed}. In addition to the ability of humans to resolve conflict, technologies may enable them to resolve conflict automatically \cite{xiao2019a3id}. Some works focus on conflict resolution considering \emph{preference aggregation} strategies, and they estimate preferences from previous service usage history \cite{cao2018attentive, chaki2021adaptive, guo2020group}. They did not take into account the rationality of \emph{interactions} and the \emph{fairness} of the residents. Hence, these aggregation strategies, which are unlikely to find out the best resolution that most residents can accept, may lead to unsatisfying service provision.

We propose a novel conflict resolution approach that integrates \emph{current service requirements} (i.e., interactions) with \emph{preferences} from previous service usages. Integrating interactions with preference is challenging due to the \emph{dynamic nature} of the residents' \emph{desires} and \emph{requirements}. For example, residents may have different requirements at different times on different days. This is why we design a \emph{preference extraction} model using the concept of \emph{temporal proximity}. We further design a \emph{preference aggregation} model using a matrix factorization-based approach, namely, \emph{Singular Value Decomposition (SVD)}. When the residents' preferences conflict heavily, we smooth their preferences by low-rank matrix factorization to ensure fairness. The concept of \emph{current resident item matrix} and \emph{ideal resident item matrix} are introduced to cater to the conflict resolution framework. The contribution of this paper is threefold:

\begin{itemize}[nosep]
    \item A novel preference extraction model using the temporal proximity concept that estimates preference scores based on previous service usage records.
    \item A novel preference aggregation model using SVD technique, current resident item matrix, ideal resident item matrix that integrates current requirements and previous preferences to find out the best item for conflict resolution.
    \item Experimental evaluation is conducted on real-world datasets to exhibit the effectiveness of the proposed framework.
\end{itemize}

%Some prefrences are transient, and some are permanent, i.e., time dependent, which is the main reason for using temporal proximity technique.

%The rest of the paper is structured as follows. Section 2 illustrates two motivation scenarios to explain the need for integrating current service requirements and previous service interactions in conflict resolution. In section 3, we introduce a set of terminologies and concepts that are used to formulate the problem. Section 4 describes the proposed conflict resolution framework. Section 5 presents our experiments to evaluate the proposed approach. Section 6 summarizes the related work on conflict detection and resolution. Section 7 discusses the constraints of the proposed framework and concludes the paper with future work.

%\vspace{-2mm}

\section{Motivation Scenario}
\begin{figure}[htbp]
\center
\includegraphics[width=\columnwidth,height=5cm,keepaspectratio]{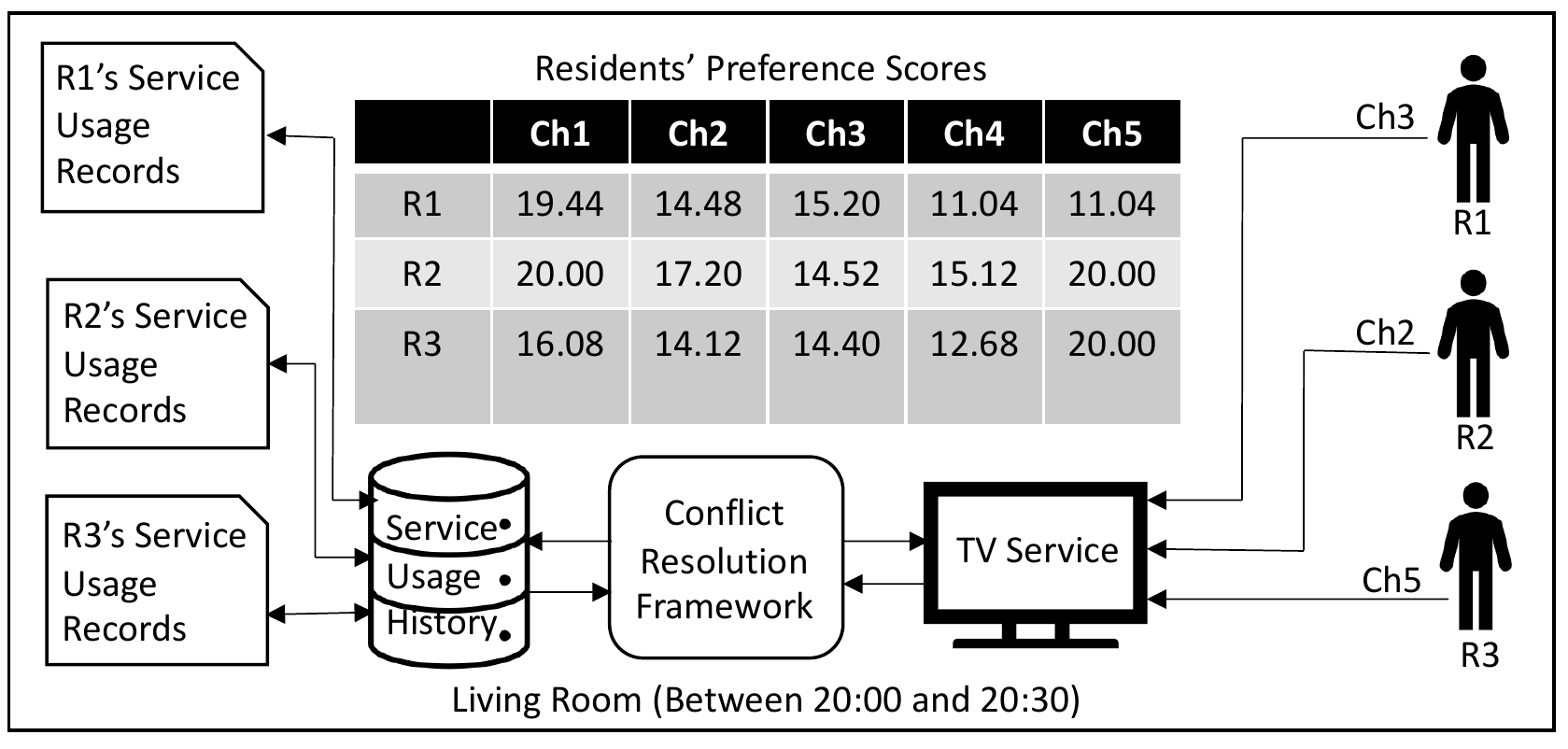}
\vspace{-2mm}
\caption{Residents' current requirements and preferences from previous usage.}
\vspace{2mm}
\label{motivation}
\end{figure}

%\vspace{-4mm}
%We consider the following two motivating scenarios to signify the need of integrating current requirements with past interactions for conflict resolution.
We consider the following motivating scenario to demonstrate the significance of our work.
%\emph{Scenario 1:}
Suppose three residents (R1, R2, R3) want to watch TV between 20:00 and 20:30 in the living room. However, they have different channel requirements. R1, R2, R3 want to watch channels Ch3, Ch2, and Ch5, respectively (Fig. \ref{motivation}). A conflict occurs since the TV cannot telecast more than one channel simultaneously (assuming the TV does not have multi-screen/split-screen features). Note that, channel is a functional property of a TV service. In this case, action may be taken to eliminate the conflict. The system may: (i) select a channel based on priorities (i.e., residents' can be prioritized based on age and/or role in a family) \cite{nurgaliyev2017improved}, (ii) adopt the use first strategy (i.e., whoever wants to use the TV first, only his/her preferred channel will be telecast) \cite{lee2019situation}, (iii) randomly pick a channel, (iv) inform users that they should explicitly resolve the conflict. However, the best choice according to these selections may still leave some residents feeling dissatisfied and slighted. Moreover, unresolved or inadequately resolved conflicts tend to result in tension, which may trigger or intensify posterior conflicts. The objective of conflict resolution is to offer a smoother and more pleasant user experience. In this regard, there is a need to have a methodology that incorporates residents' intentions (i.e., current requirements) and preferences (i.e., prior interactions). We aim to maximize residents' satisfaction by providing services that may be preferred by the majority of them. Prior interactions uncover information such as hidden patterns, correlations, habits, and preferences.

%\vspace{-1mm}

%\vspace{-2mm}

Let us assume, we know the preference scores of each channel of the residents (table in Fig. \ref{motivation}). The preference scores are calculated based on the residents' prior service interactions. The procedures of computing preference scores are showed in the proposed framework (section 4.2.1). Preference aggregation methods such as \emph{average (AVG)}, \emph{least-misery (LM)}, and \emph{most-pleasure (MP)} can be used to select the preferred channel \cite{cao2018attentive, guo2020group}. However, these methods cannot always generate a fair solution for each member in a group, leading to low satisfaction.

\vspace{2mm}
\begin{table}[htbp]
%\vspace{-5mm}
\caption{Results of preference aggregation methods}
\vspace{-6mm}
\label{motivationTable}
\center
\begin{tabular}{|c|c|c|c|c|c|}
\hline
\textbf{Methods} & \textbf{Ch1} & \textbf{Ch2} & \textbf{Ch3} & \textbf{Ch4} & \textbf{Ch5}\\
\hline
AVG & 18.51 & 15.27 & 14.71 & 12.95 & 17.01\\ \hline
LM & 16.08 & 14.12 & 15.20 & 11.04 & 11.04\\ \hline
MP & 20.00 & 17.20 & 15.20 & 15.12 & 20.00\\
\hline
\end{tabular}
%\vspace{-2mm}
\end{table}

%only aggregating preference scores without considering the current requirements may not give the best solution. 
%Then, we identify this as a preference aggregation task mainly applied to group recommendations.

%In group recommendation, preference aggregation is classified into two categories: (i) preference aggregation (PA) approaches and (ii) score aggregation (SA) approaches\cite{cao2018attentive, guo2020group}. PA approaches first aggregate group members' profiles into one profile and then make recommendations based on the aggregated profile. On the contrary, SA approaches, at first, recommend items for each group member and then aggregate the final recommendation for the group. Three methods, such as \emph{average (AVG)}, \emph{least-misery (LM)}, and \emph{most-pleasure (MP)}, are used in almost all the previous studies \cite{cao2018attentive, guo2020group}. AVG method averages the group member's preference scores as the group preference score since it assumes that each member has an equal contribution to the final group decision. LM method adopts the minimum score of individuals as the group preference score by assuming that the least satisfied member determines the final group decision. MP method, by contrast, uses the maximum score of individuals as the group preference score as it assumes that the most satisfied member determines the final group decision. Nevertheless, these methods cannot always generate a fair solution for each member in a group, leading to low satisfaction.

We apply these methods to the preference table mentioned in Fig. \ref{motivation} and get the results (Table \ref{motivationTable}). The AVG method selects the channels with the highest average ratings, Ch1 and Ch5 (if we consider the top two items). LM method selects Ch1 and Ch3 whereas MP method selects Ch1 and Ch5. We observe that both AVG and MP selects Ch5. Though R2 and R3 have a high preference for Ch5, R1 has a relatively low preference score. Ch5 is an unfair recommendation for R1. Note that, all the methods select Ch1 which is not a suitable selection, because none of the residents requests this channel in the current situation. 
%Conversely, Ch2's preference scores are close and relatively higher than Ch3's. Thus, Ch2 may be selected considering the current situation.

%The average preference scores for Ch1, Ch2, Ch3, Ch4, Ch5 are 18.51, 15.27, 14.71, 12.95, 17.01, respectively.  The lowest preference scores for Ch1 to Ch5 are 16.08, 14.12, 15.20, 11.04, 11.04, respectively. The highest preference scores for Ch1 to Ch5 are 20.00, 17.20, 15.20, 15.12, 20.00, respectively. This is not a suitable recommendation considering the current situation.

% \emph{Scenario 2}:
% Suppose two residents (R1, R2) are trying to use a light service simultaneously at the same location. R1's requirement is 200 lumens and R2's requirement is 800 lumens. A conflict occurs since the light cannot satisfy the requirements of these two residents. Note that, luminous intensity is the non-functional property of a light service. The simple approach to resolving this conflict is calculating the average value of these two requirements and setting the light luminosity at 500 lumens. In practice, some residents may have flexible preferences (i.e., accommodating) whereas some may have inflexible preferences (i.e., strict). In this context, assume that R1 has flexible preferences (i.e., 200-600 lumens) and R2 has strict preferences (i.e., 600-800 lumens). Setting light luminosity at 500 lumens may create dissatisfaction to R2. The light illumination may be set close to 600 lumens so that it falls into the range of both R1 and R2's preferences. Preference range can be computed from the previous usage records.
 
Hence, conflict resolution is situation-specific and dynamic. There is a need for a conflict resolution framework that integrates current requirements with previous usage patterns to extract preferences. The objective of conflict resolution is to enhance the residents’ overall satisfaction when a conflict occurs.
%All the existing methods make recommendations based on the previous preferences; they do not consider the current requirements.

%\vspace{-2mm}
\section{Preliminaries and Problem Formulation}
%\vspace{-2mm}
We represent the notion of \textit{IoT service}, \textit{IoT service event} and \textit{IoT service request} to explain the concept of \textit{IoT service conflict}. The definitions of IoT service, IoT service event and IoT service request have been adopted from \cite{chaki2020conflict}. 
% We focus on \textit{shareable} IoT services where conflicts may arise. A \textit{shareable} IoT service serves multiple users at the same time and location. Radio, television, DVD, AC, light, heater, and fan are some examples of shared IoT services. A \textit{non-shareable} IoT service serves only one user at a time. Examples are toaster, microwave oven, electric kettle, and washing machine.

An \textit{IoT Service ($S$)}, is a tuple of \big \langle \textit{$S_{id}$, $S_{name}$, $F$, $Q$}\big \rangle \hspace{0.15 cm}where:
%\vspace{-2mm}
\begin{itemize}[nosep]
  \item \textit{$S_{id}$} represents the unique service identifier (ID).
  \item \textit{$S_{name}$} is the name of the service.
  \item \textit{$F$} is a set of \big \{\textit{$f_1$,$f_2$,...,$f_n$}\big \} where each $f_i$ is a functional attribute of a service. The purpose of having a service is considered as the function of a service.
  \item \textit{$Q$} is a set of \big \{\textit{$q_1$,$q_2$,...,$q_m$}\big \} where $q_j$ is a non-functional attribute of a service.
\end{itemize} 

%\vspace{-1mm}

%For example, a TV service is represented as \big\langle\textit{2, TV, \{telecasting programs, receptor for security camera\}, \{\$1000, 3 years, 500 watts}\big\rangle. 2 is a unique id and TV is the name of a service. Functional properties include telecasting programs via channels (e.g., Fox, Discovery), and a TV may act as a receptor for a security camera. \{\$1000, 3 years, 500 watts\} is a set of non-functional properties like price, warranty, and electricity consumption rate, respectively.

An \textit{IoT Service Event ($SE$)} records the service state along with its user, execution time and location during the service manifestation (i.e., turn on, turn off, increase, decrease, open, close). An \textit{IoT Service Event Sequences ($SES$)} is a set of \big \{\textit{$SE_1$, $SE_2$, $SE_3$,.......$SE_k$}\big\} where each $SE_i$ is a service event. Occupants usually interact with IoT services for various household chores and the \textit{previous} interactions are recorded as IoT service event sequences. An IoT service event is a tuple of \big \langle \textit{$SE_{id}$, \{$S_{id}, F, Q\}, T, L, U$}\big \rangle \hspace{0.15 cm} where:

%\vspace{-2mm}

\begin{itemize}[nosep]
  \item \textit{$SE_{id}$} is the unique service event ID.
  \item \textit{$S_{id}$} is a unique ID of the enacted service. \textit{$F$} is a set of functional attributes. \textit{$Q$} is a set of non-functional attributes. 
  %\item \textit{$F$} is a functional attribute of the enacted service.
  %\item \textit{$Q$} is a non-functional attribute of the enacted service.
  \item \textit{$T$} is the time interval of the service consumption. $T$ is a tuple of \big \langle $SET_s, SET_e$\big \rangle \hspace{0.15 cm}where $SET_s$ and $SET_e$ represent the start time and end time of the service.
  %consumption, respectively.
  \item \textit{$L$} is the service event location and \textit{$U$} is user who consumed the service.
\end{itemize}

%\vspace{-1mm}

%An example of an IoT service event is \big\langle \textit{56, \{2, MTV, 80 dB, \{2019-03-14T19:30, 2019-03-14T20:30\}, bedroom, U3}\big\rangle. 56 is the ID of this service event and 2 is the ID of the TV. The user watched MTV channel keeping 80 dB volume on the TV between 19:30 and 20:30 on March 14, 2019. Bedroom is the location and U3 is the user who consumed this service.

An \textit{IoT Service Request ($SR$)}, is an instantiation of a service and it represents a resident's \textit{current} service requirement. An \textit{IoT Service Request Sequences ($SRS$)} is a set of \big \{\textit{$SR_1$, $SR_2$, $SR_3$,.......$SR_n$}\big\} where each $SR_i$ is an IoT service request. Residents' current service requirements are recorded as IoT service request sequences. An IoT Service Request ($SR$) is a tuple of \big \langle \textit{$SR_{id}$, \{$S_{id}, F, Q\}, \{SRT_s, SRT_e\}, L, U$}\big \rangle \hspace{0.15 cm} where:

\begin{itemize}[nosep]
  \item \textit{$SR_{id}$} is the unique service request ID.
  \item \textit{$S_{id}$} is a unique ID of the requested service. \textit{$F$} is a functional attribute and \textit{$Q$} is a non-functional attribute of the requested service.
  \item \{\textit{$SRT_s$,$SRT_e$}\} represent the requested service's start time and end time.
  \item \textit{$L$} is the location of the service and \textit{$U$} is the user of the service.
\end{itemize}

%\vspace{-2mm}

%An example of an IoT service request is \big\langle \textit{14, \{2, Discovery, 50 dB, \{19:30, 20:30\}, living room, U1}\big\rangle. 14 is the ID of this service request and 2 is the ID of the TV. The user wants to watch the TV between 19:30 and 20:30 on the day when the request is made. Discovery and 50 dB are the values of functional and non-functional attributes such as channel and volume of TV. Living room and U1 denote the location and user of the requested service, respectively.

%\vspace{-6mm}

\subsection{Formal Problem Statement}
%\vspace{-2mm}
An IoT service ($S$) is associated with a set of functional and non-functional properties. An IoT service event ($SE$) illustrates a resident's previous service usage, in conjunction with time and location. IoT service event sequences ($SES$) record all the history of service events and preferences can be estimated from these previous events. An IoT service request ($SR$) captures a resident's current service usage requirement. Multiple residents' requirements are stored in service request sequences ($SRS$). A conflict may emerge since different residents may have different service requirements. Consequently, a conflict resolution ($Res$) technique is required to maximize the satisfaction of the residents. Given this information, the paper aims to identify a function $F(S, SRS, SES)$, where $Res \approx F(S, SRS, SES)$. In other words, our goal is to resolve conflict using service-related, current requirement-related and previous usage-related data.

%\vspace{-6mm}

\section{Conflict Resolution Framework}
\begin{figure}[htbp]
\center
\includegraphics[width=\columnwidth,height=6cm,keepaspectratio]{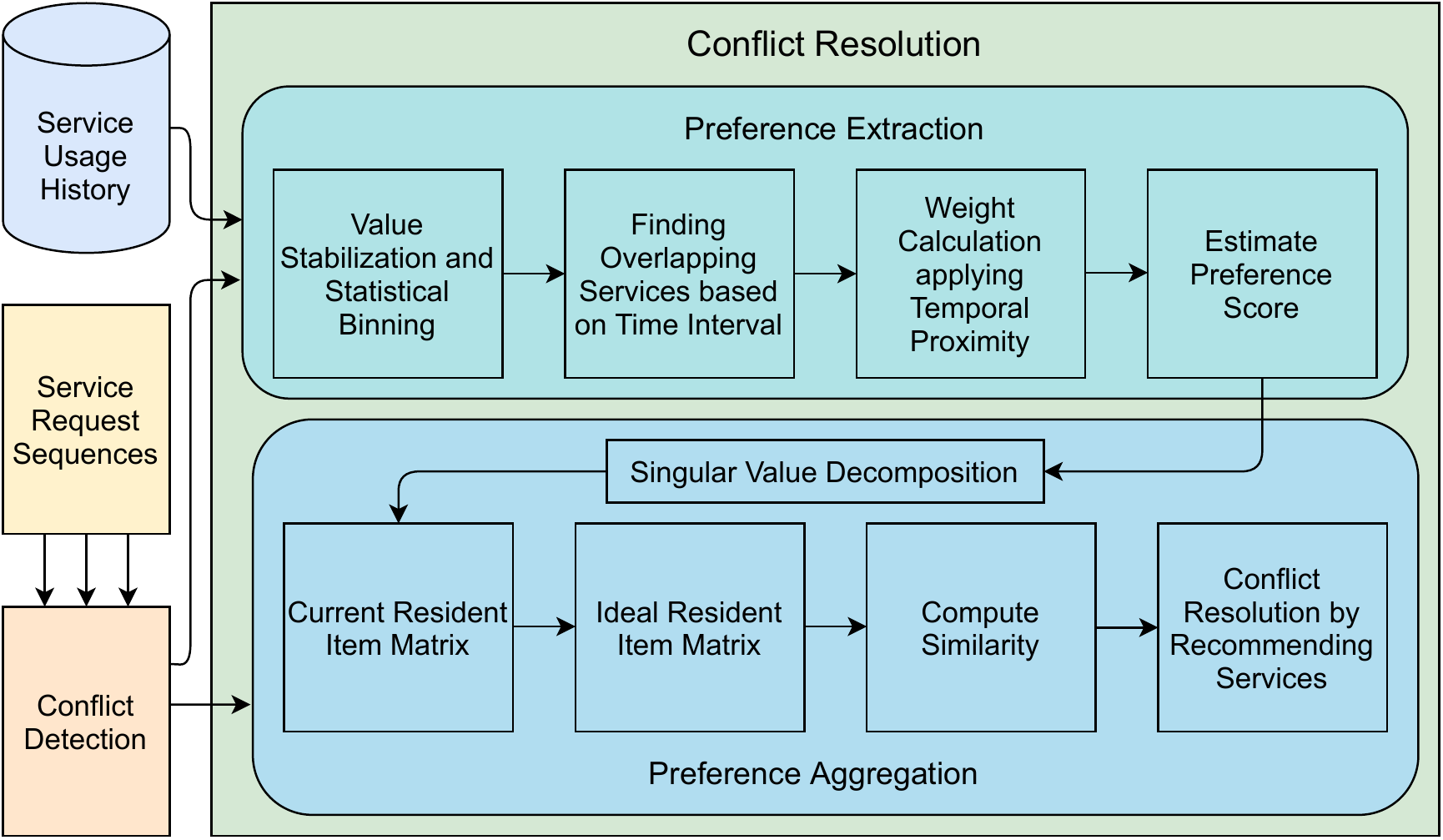}
%\vspace{-5mm}
\caption{IoT service conflict resolution framework.}
\vspace{-2mm}
\label{framework}
\end{figure}

%\vspace{-4mm}
The proposed conflict resolution framework has 4 modules: (i) service event sequences (a.k.a., service usage history), (ii) service request sequences, (iii) conflict detection, and (iv) conflict resolution (Fig. \ref{framework}). Service usage history and service request sequences modules are described in Section 3. In this section, we thoroughly describe conflict detection and conflict resolution modules.

%\vspace{-6mm}

% \begin{figure}
% \includegraphics[width=\textwidth]{fig1.eps}
% \caption{A figure caption is always placed below the illustration.
% Please note that short captions are centered, while long ones are
% justified by the macro package automatically.} \label{fig1}
% \end{figure}
%\vspace{-1mm}

%\subsection{Service Event Sequences}
%\vspace{-1mm}

%\vspace{-4mm}

\subsection{Conflict Detection}
%\vspace{-2mm}
Conflict detection is the pre-requisite of conflict resolution. An \textit{IoT service conflict} occurs when a service cannot satisfy the requirements of multiple users at the same time and location. Conflicts are defined considering the current requirements of occupants, and these requirements are generated from the IoT service requests. Given two service requests ($SR_i$, $SR_j$), the following conditions have to be satisfied to be considered as a conflict situation.

%Due to lack of space, we only provide the definitions of different types of conflicts.
\begin{itemize}[nosep]
    \item $L_{S_i}$ $\simeq$ $L_{S_j}$, meaning, two services ($S_i$, $S_j$) are executed at the same location.
    \item $(SRT_{s_i}, SRT_{e_i}) \cap (SRT_{s_j}, SRT_{e_j})) \neq \emptyset$, denoting that two service requests ($SR_i$, $SR_j$) are invoked at the same time and there is a temporal overlap.
    \item $U_{S_i} \neq U_{S_j}$, meaning, these two requests are invoked by two different users.
    \item $\exists Q_k \in S.Q: S_i.Q_k \neq S_j.Q_k$; there exist at least one property which is different between $S_i.Q$ and $S_j.Q$.
\end{itemize}

%\vspace{-2mm}

We adopt the conflict detection algorithm proposed in \cite{chaki2020conflict}. This component is not the core of our contributions; however, it produces the input for the conflict resolution module, which holds the present work's core contributions.

%\vspace{-6mm}

\subsection{Conflict Resolution}
%\vspace{-3mm}
Conflict resolution is conducted in two phases: (i) preference extraction and (ii) preference aggregation. Phase 1 mines previous service usage records and extracts occupants' preferences. Phase 2 aggregates all the occupants' preferences and selects the service that may give relatively high satisfaction to them.

%\vspace{-6mm}

\subsubsection{4.2.1   Preference Extraction:}
In this phase, we estimate users' preferences for a service based on previous usage records. Conflict detection module outputs the name of the conflicting services and the overlapping time-period where a conflict occurs. These are the inputs of this phase along with previous service usage history. It extracts residents' service usage patterns from the previous history. Then it computes the preference score of frequently used services. 

%\vspace{-5mm}

\paragraph{\textbf{\emph{Value stabilization and statistical binning.}}}

Some service event ($SE$) data need to go through some pre-processing steps such as \textit{value stabilization} and \textit{statistical binning} \cite{mishra2020alternate}. Several values are advertised within a short period of time for some service attributes, where only the final value is relevant. For example, browsing through TV channels before settling down at a final channel. In this work, we only consider the final settled down value while measuring the service usage preference of the residents. We compute the preference score of each attribute based on categorical values. However, there are some attributes that have numerical values. Therefore, we apply a statistical method called data binning. It takes the continuous numerical values and puts them into multiple categories. We use a dynamic programming approach to get the optimal bin \cite{mishra2020alternate}. 
%At the end of this step, all the $SE$ of all residents are stored in a database ($DB$).

%\vspace{-5mm}

\paragraph{\textbf{\emph{Finding overlapping service events.}}}

This step scans the previous history to find out all the overlapping service events (algorithm 1). The input of this algorithm is the previous service usage dataset ($DB$) and the conflicting time-period. All the previous events that have overlap with the given conflicting time interval are the output of this algorithm. For example, a conflict related to a TV service occurs in the living room between 20:00 and 20:30. This component searches all the TV service events which previously occurred, either partially or fully, between 20:00 and 20:30 in the living room; stores them into a list ($OSE$). This list contains the overlapping service events along with their timestamps.
%because of two different channel requirements from two residents. This component searches all the TV service events which occurred previously between 20:00 and 20:30 in the living room (lines [11-19] in algorithm 1). Before that, it clusters services location wise. A smart home typically has locations such as a kitchen, a bedroom, a living room, and a bathroom. Since each service is located in a search space, each service event is associated with a search space. We cluster service events ($SE$) based on location ($L$) (lines [2-10] in algorithm 1). Suppose a DVD and a TV are located in a living room. For the living room, $SE$ cluster is represented as $LSE_{living} = \big \langle SE_{DVD}, SE_{TV} \big \rangle$.

\setlength{\textfloatsep}{0pt} % For vertically reducing space under algorithm.
\begin{algorithm}[t!]
\small
\caption{Overlapping Service Events}\label{alg:algorithm1}
\begin{algorithmic}[1]
\REQUIRE
$DB$, $[s,e]$ // conflicting time-period $[s,e]$
\ENSURE
$OSE$ // overlapping service events along with time interval

\STATE {$TM = \emptyset, OSE = \emptyset$}

% \item[] // Clustering services based on location
% \FOR{\textbf{each} $se_i$ in $DB$}
% \FOR{\textbf{each} $l_j$ in $se.l$}
% \FOR{\textbf{each} $s_k$ in $se.S$}
% \IF{$s_k.l$ is equal to $l_j$}
% \STATE $LSE_i \leftarrow insert(s_k)$
% \ENDIF
% %\STATE $us_i \leftarrow unique(services)$
% \ENDFOR
% \ENDFOR
% \ENDFOR
\item[] // Finding overlapping service events
%\textbf{Step 2. Find Common Services}
\FOR{\textbf{each} $se_i$ in $DB$}
\FOR{\textbf{each} $s_j$ in $se_i$}
%\STATE $CS \leftarrow count(s_j)$ 
\IF{$s_j.L == se_i.L$}
\IF{$s_j.SET_s$ or $s_j.SET_e$ falls between $[s,e]$}
\STATE $TM \leftarrow addTimeInterval(s_j.SET_s, s_j.SET_e)$ 
\STATE $OSE \leftarrow insert(s_j, TM)$
\ENDIF
\ENDIF
\ENDFOR
\ENDFOR
%\STATE $SV \leftarrow sort(TM)$
%\STATE $SS \leftarrow top (SV,k)$
%\STATE $TI \leftarrow timeInterval(S)$
%\STATE $OSE \leftarrow overlap(SV)$
\RETURN $OSE$

\end{algorithmic}
\end{algorithm}

%\vspace{-5mm}

\paragraph{\textbf{\emph{Weight calculation applying temporal proximity technique.}}}

%Some service events overlap for a small amount of time. For instance, one day, a resident watched TV between 20:00 and 20:30 pm. Another day they watched TV between 20:25 pm and 21:30 pm. The overlapping period is minimal (only 5 minutes, i.e., between 20:25 pm and 20:30 pm). 
We use temporal proximity strategy to find out weight of the relevant events. Temporal proximity technique for evaluating the distance between time-interval data is adopted from \cite{shao2016clustering}. For each service event, $SE_i$, we use a function $f_i$ with respect to $t$ to map the temporal aspect of $SE_i$. Event start time and end time are represented with $SE_{i_{st}}$ and $SE_{i_{et}}$, respectively. $f_i$ is formalized in Equation \eqref{proximity}.

\vspace{-2mm}

\begin{equation}
\label{proximity}
%\vspace{-2mm}
    f_i(t) = 
    \begin{cases}
      1, & t \in [SE_{i_{st}}, SE_{i_{et}}] \\
      0, & otherwise
    \end{cases}
    %\vspace{-1mm}
\end{equation}

We generate a set of functions ${f_1, f_2, ... f_n}$ corresponding to the service event instances ($SE$). Equation \eqref{temporal proximity} calculates the temporal proximity ($temp_{prox}$) for all the overlapping events.

\vspace{-4mm}

\begin{equation}
%\vspace{-2mm}
    temp_{prox} = \frac{\int_{t_1}^{t_{2n}}\sum_{i=1}^{n}f_i(t)dt}{(t_{2n}-t_1).n}
    \label{temporal proximity}
    %\vspace{-1mm}
\end{equation}

Here, $t_1$ and $t_{2n}$ are the first and last time information of overlapped events from $OSE$, and $n$ is the number of instances. Consider the following two events of watching TV from a resident, R1. One Sunday, they watched TV between 20:00 and 21:00; another Sunday, they watched TV between 20:45 and 21:45. Using Equation \ref{temporal proximity}, the temporal proximity of these two events can be calculated as ((20:45-20:00)+(21:00-20:45)*2+(21:45-21:00))/((21:45-20:00)*2=0.57. Consider another scenario where a resident, R2, watched TV between 18:00 and 19:00. Another day, they watched TV between 18:10 and 19:10. The temporal proximity of these two events can be calculated as ((18:10-18:00)+(19:00-18:10)*2+(19:00-19:10))/((19:10-18:00)*2)=0.86. Thus the latter case has more weight while calculating the preference score of watching TV service.
%The latter scenario has higher temporal proximity than the former one.

%\vspace{-5mm}

\paragraph{\textbf{\emph{Estimate preference score.}}}

We mine out the frequent service usage records and calculate the preference score for each resident. For example, if a resident watched Discovery 6 times and Fox 4 times between 20:00 and 20:30 (fractional overlapping time is also considered while calculating frequency) on the last 10 days, then, the frequency of each channel for this resident would be \big\langle \textit{\{Fox, 4\}, \{Discovery, 6\}}\big\rangle. Frequency ($F$) is formally defined in Equation \ref{freq} and \ref{frequency} \cite{roushan2014university}.

\vspace{-3mm}

\begin{equation}
%\vspace{-1mm}
    S = (s_1,s_2,.....,s_n):s_i \in A
    \label{freq}
    \vspace{-2mm}
\end{equation}
%\vspace{-5mm}
\begin{equation}
%\vspace{-1mm}
    F(a) = \sum_{i=1}^{n}[s_i = a]
    \label{frequency}
    \vspace{-1mm}
\end{equation}

where the sequence $S$ contains elements of the set $A$. The frequency value $F(a)$ for an element $a$ is defined as the number of its occurrences in the sequence $S$. Then, we compute the preference score ($PS$) for each element ($a$) by multiplying the frequency value and temporal proximity as follows:

\vspace{-2mm}
\begin{equation}
    PS = \sum_{i=1}^{n}(temp_{prox}(a) \times F(a))
    %temp_{prox} = \frac{\int_{t_1}^{t_{2n}}\sum_{i=1}^{n}f_i(t)dt}{(t_{2n}-t_1).n}
    \label{ps}
    \vspace{-1mm}
\end{equation}

Let us consider the motivation scenario again. A conflict related to a TV service occurs between 20:00 and 20:30. We calculate the preference score of each resident for each TV channel from the previous usage record. Suppose in the last 100 days, Resident, R1, watched channel, Ch1, 19 times (frequency = 19) at the same time period between 20:00 and 20:30 (i.e., temporal proximity = 1). One time (frequency = 1), they watched Ch1 at the time period which partially overlaps with the current conflicting period (assume, temporal proximity = 0.44). Then, R1's preference score for Ch1 becomes (19*1)+(1*0.44)=19.44.

%\vspace{-5mm}

\subsubsection{4.2.2   Preference Aggregation:}

In this phase, we use the example from the motivation scenario 1 to illustrate preference aggregation methodology. We first create a Historical Resident Item Matrix (H) considering the highest preference score of each resident in a group. Then, we do Singular Value Decomposition (SVD) on the H matrix and construct a Current Resident Item Matrix (CRIM). CRIM deduces the features of the ideal item in latent factor space by incorporating the current requests (current requests are also represented in a matrix) \cite{hasan2018novel}. 
%Thus, dimensionality can be reduced by discarding insignificant factors to smooth the group members' preferences. 
After that, we define an Ideal Resident Item Matrix (IRIM) based on CRIM and represent the ideal item of the conflicting group in preference space. Finally, we can resolve conflicts by offering the ideal items (in this case, TV channels) that are more likely to be accepted by the residents.

%\vspace{-4mm}

\paragraph{\textbf{\emph{Current Resident Item Matrix.}}}

At first, we introduce the notion of the \emph{item set} of each group into latent space, and then we represent the current resident item matrix. Given a group of residents $G=(R_1,R_2,...,R_{|G|})$, their item set($IS$) would be $IS = (I_1,I_2,...,I_{|IS|})=\bigcup_{i=1}^{|G|}\bigcup_{j=1}^{|N|}$. Here, $IS$ represents the group's item set and each item belongs to at least one of the resident's item sets. For example, if we pick 3 items with the highest preference scores from Ch1 to Ch5 (see motivation scenario) for each resident, then, group's item set would be, $IS = (I_1 \cup I_2 \cup I_3) = \{Ch1,Ch2,Ch3,Ch5\}$ where $I_1=(Ch1,Ch3,Ch2)$, $I_2=(Ch1,Ch5,Ch2)$, $I_3=(Ch5,Ch1,Ch2)$. The union operation on $I_1,I_2,I_3$ gives $\{Ch1,Ch2,Ch3,Ch5\}$. We create the H matrix using $IS$ as follows:

\vspace{-3mm}
\begin{equation}
    H = (PS(i,j))_{|G|\times|IS|}
    \label{hrimeq}
    %\vspace{-2mm}
\end{equation}

where $(PS(i,j))_{|G|\times|IS|}$ is the preference score of resident $R_i$ to the $j-th$ item. For the example of the motivation scenario 1, using Equation \ref{hrimeq}, we get:

\vspace{-2mm}
  \begin{equation*}
  \label{hrim}
  H = 
  \setlength\arraycolsep{10pt}
      \begin{bmatrix}
        19.44 & 14.48 & 15.20 & 11.04\\
        20.00 & 17.20 & 14.52 & 20.00\\
        16.08 & 14.12 & 14.40 & 20.00\\
      \end{bmatrix}
  \end{equation*}

Preference scores of Ch1, Ch2, Ch3, and Ch5 are represented in the 1st, 2nd, 3rd and 4th columns, respectively. Ch4 is not considered since it does not belong to any resident's top-3 item list. We apply singular value decomposition (SVD) to the H matrix to produce a set of vectors corresponding to features in the matrix. We compute SVD of matrix H as follows:

\vspace{-4mm}
\begin{equation}
    H_{|G|\times|IS|} = A_{|G|\times|G|}D_{|G|\times|IS|}V_{|IS|\times|IS|}^T
    \label{svd}
    \vspace{-2mm}
\end{equation}

where $A$ is the resident-feature matrix, $D$ is the diagonal weight matrix, and $V$ is the item-feature matrix. Additionally, dimentionality reduction can be achieved by low-rank matrix approximation as follows:

\vspace{-7mm}
\begin{align}
    \tilde{H} & = A_{|G|\times|w|}D_{|w|\times|w|}V_{|IS|\times|w|}^T \nonumber\\
              & = \tilde{A}\tilde{D}\tilde{V}^T    
    \label{dimensionality}
\end{align}

\vspace{-4mm}

where $w=min\bigg\{w|\frac{\sum_{k=1}^{w} D(k,k)}{\sum_{k=1}^{|G|} D(k,k)} > \alpha\bigg\}$, $w$ denotes the significant features' number. Parameter $\alpha$ controls the degree of denoising or smoothness. When $\alpha$ is smaller, the smoothness becomes heavier. This process is required when there is a significant variance in the residents' preferences. We apply Equation \ref{svd} to our running example and get the singular value decomposition of $H=ADV^T$ as:

\vspace{-3mm}
  \begin{equation*}
  \label{A}
  A = 
  \setlength\arraycolsep{15pt}
      \begin{bmatrix}
        -0.5278 & -0.8206 & 0.2194\\
        -0.6320 & 0.2068 & -0.7469\\
        -0.5675 & 0.5328 & 0.6277\\
      \end{bmatrix}
  \end{equation*}
  
  \vspace{-2mm}
    \begin{equation*}
  \label{D}
  D = 
  \setlength\arraycolsep{15pt}
      \begin{bmatrix}
        57.1127 & 0 & 0 & 0\\
        0 & 6.8771 & 0 & 0\\
        0 & 0 & 1.8235 & 0\\
      \end{bmatrix}
  \end{equation*}
  
  \vspace{-2mm}
    \begin{equation*}
  \label{V}
  V = 
  \setlength\arraycolsep{15pt}
      \begin{bmatrix}
        -0.5607 & -0.4724 & -0.3176 & 0.6013\\
        -0.4644 & -0.1166 & -0.4422 & -0.7584\\
        -0.4442 & -0.2614 & 0.8385 & -0.1767\\
        -0.5221 & 0.8336 & 0.0211 & 0.1792\\
      \end{bmatrix}
      \vspace{-2mm}
  \end{equation*}

where $A(i,k)$ measures the preference of resident $R_i$ to feature $F_k$, $D_{k,k}$ denotes the feature's importance, and the preference of item $I_j$ to feature $F_k$ is measured by $V_{j,k}$. For the running example, we set $\alpha = 0.97 $ in Equation \ref{dimensionality} to denoise $D$ to $D(1:2, 1:2)$, and we get:
\vspace{-6mm}

\begin{align}
\tilde{A} &= 
\setlength\arraycolsep{5pt}
    \begin{bmatrix}
        -0.5278 & -0.8206\\
        -0.6320 & 0.2068\\
        -0.5675 & 0.5328\\
    \end{bmatrix}
&
\tilde{D} &= 
\setlength\arraycolsep{2pt}
\begin{bmatrix}
        57.1127 & 0\\
        0 & 6.8771\\
\end{bmatrix}
&
\tilde{V} &= 
\setlength\arraycolsep{7pt}
\begin{bmatrix}
         -0.5607 & -0.4724\\
        -0.4644 & -0.1166\\
        -0.4442 & -0.2614\\
        -0.5221 & 0.8336 \nonumber
\end{bmatrix}
\end{align}

%After the decomposition, we show how to aggregate the unitary preference for this group and obtain the final recommended items.

\vspace{-4mm}
Integrating the residents' preferences in the decomposed latent space with current service requests ($SR$) is defined as the current resident item matrix (CRIM). We formally define CRIM for each group as:

\vspace{-4mm}
\begin{equation}
    CRIM = \frac{1}{|G|}\sum_{i=1}^{|G|}\sum_{j=1}^{|IS|}SR_{j}^{i}\tilde{V}(IS_{j}^{i}, 1 : w)
    \label{crim}
    \vspace{-2mm}
\end{equation}

where $IS_{j}^{i}$ is the position of $R_i$'s preferred item $S_{j}^{i}$ in item set $IS$. Applying Equation \ref{crim} to the running example\footnotemark, we get current resident item matrix as:

\footnotetext {R1 requests Ch3, R2 requests Ch2, R3 requests Ch5. In the $\tilde{V}$ matrix, row1, row2, row3, and row4 represent Ch1, Ch2, Ch3, and Ch5, respectively.}

%Thus, $CRIM$ could be viewed as the ideal item in the latent factor space after preference smoothness and aggregation.

\vspace{-4mm}
\begin{align}
    CRIM & = ((1.00,0.00,0.00).\tilde{V}([3,2,4],1:2)+(1.00,0.00,0.00).\tilde{V}([2,3,4],1:2) \nonumber\\
         & + (1.00,0.00,0.00).\tilde{V}([4,2,3],1:2))/3 \nonumber\\
         & = ((-0.4442,-0.2614)+(-0.4644,-0.1166)+(-0.5221,0.8336))/3     \nonumber\\
         & = (-0.48, 0.15) \nonumber
    \label{crimexp}
\end{align}

%\vspace{-6mm}

\paragraph{\textbf{\emph{Ideal Resident Item Matrix.}}}

Given CRIM, if we want to resolve conflict (i.e., provide group-oriented optimal services), we have to figure out the most similar items to CRIM. To find similar items, we project CRIM to the first matrix $H$ by matrix multiplication. Thus, we define ideal resident item matrix ($IRIM$) as:

\vspace{-4mm}
\begin{equation}
    IRIM = \tilde{A} \times \tilde{D} \times CRIM^T
    \label{irim}
    \vspace{-2mm}
\end{equation}

$IRIM$ is the prototype of aggregated preferences in preference space and can be considered as ideal items. When a decision is made considering conflicting requirements, each element in $IRIM$ implies to what degree this resident's preference can be considered or expressed in a particular conflicting situation. Consequently, we tend to select candidate items whose preference scores are very close to the given group's $IRIM$ scores for group-oriented service. We compute the ideal resident item distance ($IRID$) to measure the similarity between the currently requested item ($RI_j$) and $IRIM$. We define $IRID(RI_j, IRIM)$ as:

\vspace{-4mm}
\begin{equation}
    IRID(RI_j, IRIM) = ||H(1:|G|, RI_j)-IRIM||_2
    \label{irid}
    \vspace{-2mm}
\end{equation}

We can then choose the most preferred items with the lowest $IRID$ values as the final selections for conflict resolution considering residents are more likely to agree on the items similar to the aggregated unitary preference. Applying Equation \ref{irim} to the running example, we get ideal resident item matrix as follows:

\vspace{-6mm}
\begin{equation*}
\begin{split}
    IRIM = \tilde{A}\times \tilde{D} \times CRIM^T &=     \begin{bmatrix}
        -0.5278 & -0.8206\\
        -0.6320 & 0.2068\\
        -0.5675 & 0.5328\\
    \end{bmatrix} \times \begin{bmatrix}
        57.11 & 0\\
        0 & 6.88\\
\end{bmatrix} \times \begin{bmatrix}
        -0.48\\
        0.15\\
\end{bmatrix} \\ 
    & = (13.623, 17.539, 16.107)^T
    \vspace{-2mm}
    \end{split}
\end{equation*}

\vspace{-3mm}

$IRIM$ is the ideal item of unitary preference in preference space for this conflicting group. The estimated items from three residents, which are more similar to $IRIM$'s corresponding elements, are better. By using Equation \ref{irid}, we get $IRID$ as:

\vspace{-6mm}
\begin{align}
    IRID(RI_1, IRIM) & = ||H(1:|G|, RI_1)-IRIM||_2 \nonumber\\
         & = ||(19.44,20.00,16.08)-(13.62, 17.53, 16.10)||_2  = 6.32   \nonumber\\
    IRID(RI_2, IRIM) & = 2.19  \;\;\;\;\;\;\;\;\;\;\;\;\;\;   IRID(RI_3, IRIM) = 3.81 \nonumber\\
    IRID(RI_4, IRIM) & = 4.93  \;\;\;\;\;\;\;\;\;\;\;\;\;\;   IRID(RI_5, IRIM) = 5.28 \nonumber
    \label{crimexp}
    \vspace{-2mm}
\end{align}

%\vspace{-2mm}

$IRID(RI_2,IRIM)$ and $IRID(RI_3,IRIM)$ are the lowest, which means Ch2 and Ch3 are similar to the ideal resident item matrix. In other words, Ch2, Ch3 are closer to the best choice of residents. If we pick Ch2 and Ch3 as conflict resolutions, it's more likely that each resident has a relatively high satisfaction. 

%\vspace{-6mm}
\section{Experimental Results and Discussion}
%\vspace{-3mm}
\subsection{Experimental Data}
%\vspace{-3mm}
We use a dataset collected from the Center for Advanced Studies in Adaptive Systems (CASAS) to evaluate the proposed conflict resolution framework \cite{cook2012casas}. We use four individual residents' service interaction records (labels HH102, HH104, HH105, HH106) and merge them to mimic the environment of multi-resident smart homes. We select these labels as they contain activities of a similar period (between June 15, 2011, and August 14, 2011). Descriptions of dataset attributes are displayed in Table \ref{dataset}. The dataset has ``Watch\_TV" activity label, however, the channel information is missing. Hence, we augment the dataset by randomly assigning channel values based on a uniform distribution. We use another dataset, namely CAMRa2011, which has 145096 ratings for 7740 movies. It has the rating records of 602 residents from 290 households \cite{cao2018attentive}. Among these 290 households, 272 households have 2 residents, 14 households have 3 residents, and 4 households have 4 residents. The rating scale is [1-100]. We consider the rating score as the preference score and each movie as a TV channel to evaluate our proposed framework. Since this dataset does not have any timestamps, we randomly generate the timestamp records based on a uniform distribution.

\vspace{4mm}

\begin{table}[htbp]
%\vspace{-5mm}
\caption{Description of the dataset attributes}
\vspace{-4mm}
\label{dataset}
\center
\begin{tabular}{|c|l|p{6.3cm}}
\hline
\textbf{Attributes} &  \textbf{Description} \\
\hline
Date &  The service execution date\\ \hline
Time &  The service execution time\\ \hline
Sensor & \parbox{10cm}{Name of the sensors such as motion sensors, light switch, light sensors, door sensors, temperature sensors}\\\hline
Status & ON, when the service starts, and OFF, when the service stops\\
\hline
\end{tabular}
%\vspace{-2mm}
\end{table}

%\vspace{-2mm}

%\vspace{-4mm}

\subsection{Experimental Setup}

%\vspace{-2mm}

In the experiments, we mainly evaluate the \emph{preference aggregation} model. We did not find any relevant work to compare the \emph{preference extraction} model. This paper is the first attempt to extract preferences from prior service interactions, aiming to compute preference scores for the purpose of conflict resolution.The evaluation of preference aggregation is not affected by preference extraction since all the aggregating strategies are implemented in the same settings of preference scores. The $\alpha$ parameter in SVD is set as 0.97 without special illustration. 
%For each data set, the data is divided into 5 parts, and we adopt the 5-fold cross-validation. Each time 4 parts  are selected as training set, and the rest one is used for test.

%\vspace{-3mm}

\subsubsection{5.2.1  Experimental Methods:} We select three state-of-the-art preference aggregation methods used on group recommendation as baselines. They are average (AVG) strategy, least-misery (LM) strategy, and most-pleasure (MP) strategy.

%\vspace{-3mm}

\subsubsection{5.2.2  Metrics:} We recommend items for conflict resolution. Here, items refer to the values of service attributes. Two widely used group recommendation metrics are utilized for the evaluation of the proposed model. We calculate the average value of all our results on these metrics in all the conducted experiments.

%\vspace{-3mm}

\paragraph{\textbf{\emph{Satisfaction Gain (SG).}}} SG metric measures the satisfaction of a group to a list of recommended items \cite{shahabi2003adaptive}. $SG=\frac{1}{|G|}\sum_{j=1}^{|G|}\sum_{k=1}^{|L|}PS(j,k)$,
where $|G|$ represents the group, $|L|$ denotes the recommended items, $PS(j,k)$ is the preference score of each member on item ($I_k$) and $I_k$ is an adopted item. Adopted items refer to the items that have been used more than 60\% times by the residents.

%\vspace{-3mm}

\paragraph{\textbf{\emph{Harmonic (H).}}} H metric estimates the equity of the recommended items to the group, $H=|G|/(\sum_{j=1}^{|G|}\frac{1}{\sum_{k=1}^{|L|}R(j,k)})$. If the value of harmonic metric is high, it can be said that the recommendation is fair to all members \cite{carvalho2013users}.

%\vspace{-4mm}

\subsection{Experimental Results}
%\vspace{-2mm}
\subsubsection{5.3.1  Efficiency Results:}
The efficiency results are illustrated by comparing different methods and their running times. The average running times (in seconds) of each method on CASAS and CAMRa2011 datasets are displayed in Table \ref{efficiency}. These time records do not include the runtime of the preference extraction step; they include the runtime of the preference aggregation step and conflict resolution step. AVG, LM, and MP are very efficient in terms of runtime. For each group, they directly calculate item scores from the preference table. AVG, LM, and MP methods require more time on CAMRa2011 dataset than on CASAS dataset since items are denser on CAMRa2011. Our approach takes a long time on both datasets than these methods because we compute matrix approximation for all the candidate items for each conflict situation.

\vspace{3mm}
\begin{table}[htbp]
%\vspace{-5mm}
\caption{Efficiency results (average running time in seconds)}
\vspace{-4mm}
\label{efficiency}
\center
\begin{tabular}{|c|c|c|c|c|}
\hline
\textbf{Datasets} & \textbf{AVG} & \textbf{LM} & \textbf{MP} & \textbf{Our Approach}\\
\hline
CASAS & 1.05 & 1.39 & 1.26 & 2.52\\ \hline
CAMRa2011 & 1.53 & 2.24 & 2.34 & 4.36\\
\hline
\end{tabular}
\vspace{-3mm}
\end{table}

\subsubsection{5.3.2  Effectiveness Results:}
\begin{figure}%
    \centering
    \subfloat[\centering Satisfaction gain]{{\includegraphics[width=5.5cm,height=3.5cm]{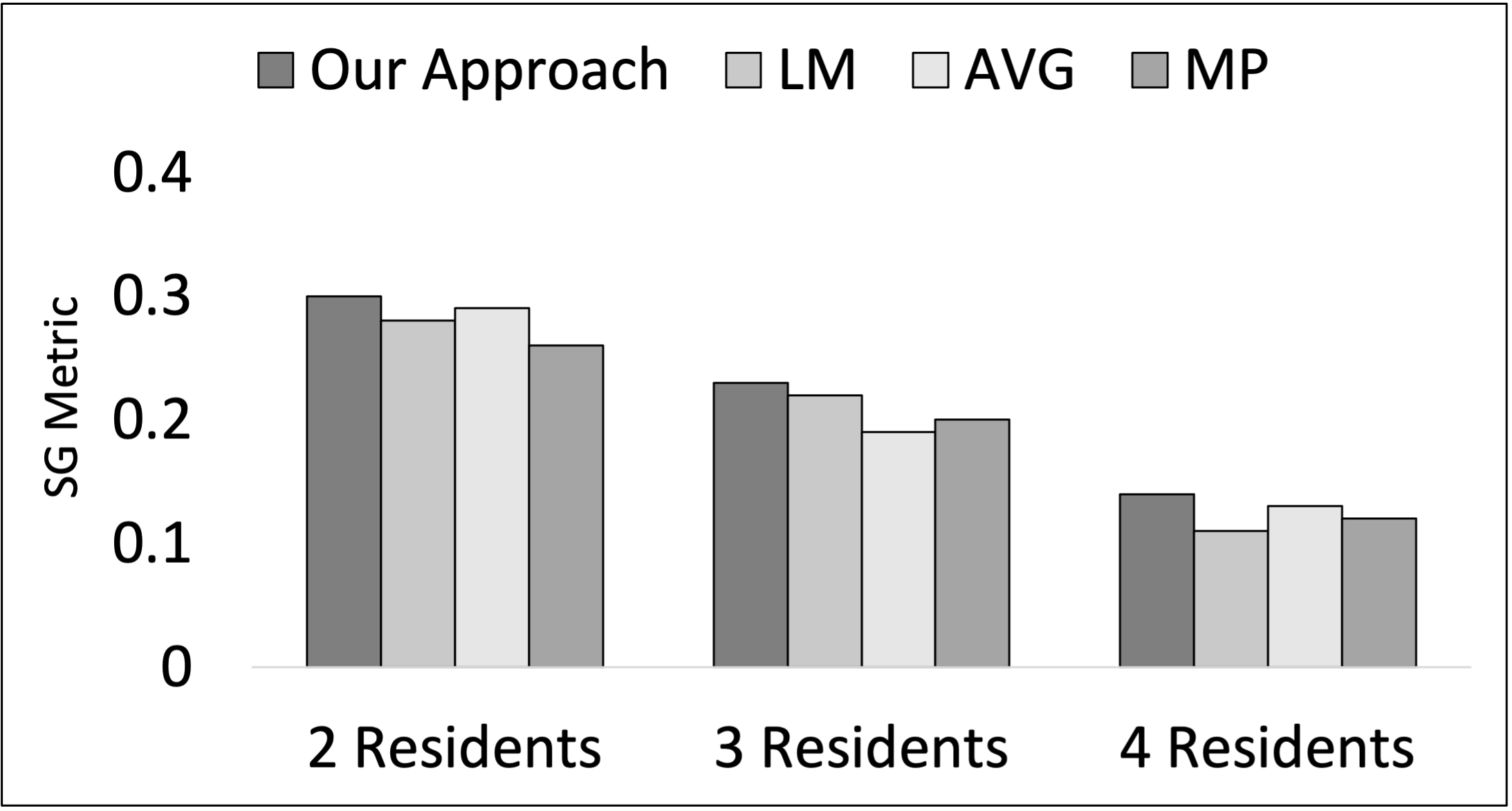} }}%
    \qquad
    \subfloat[\centering Harmonic ]{{\includegraphics[width=5.5cm,height=3.5cm]{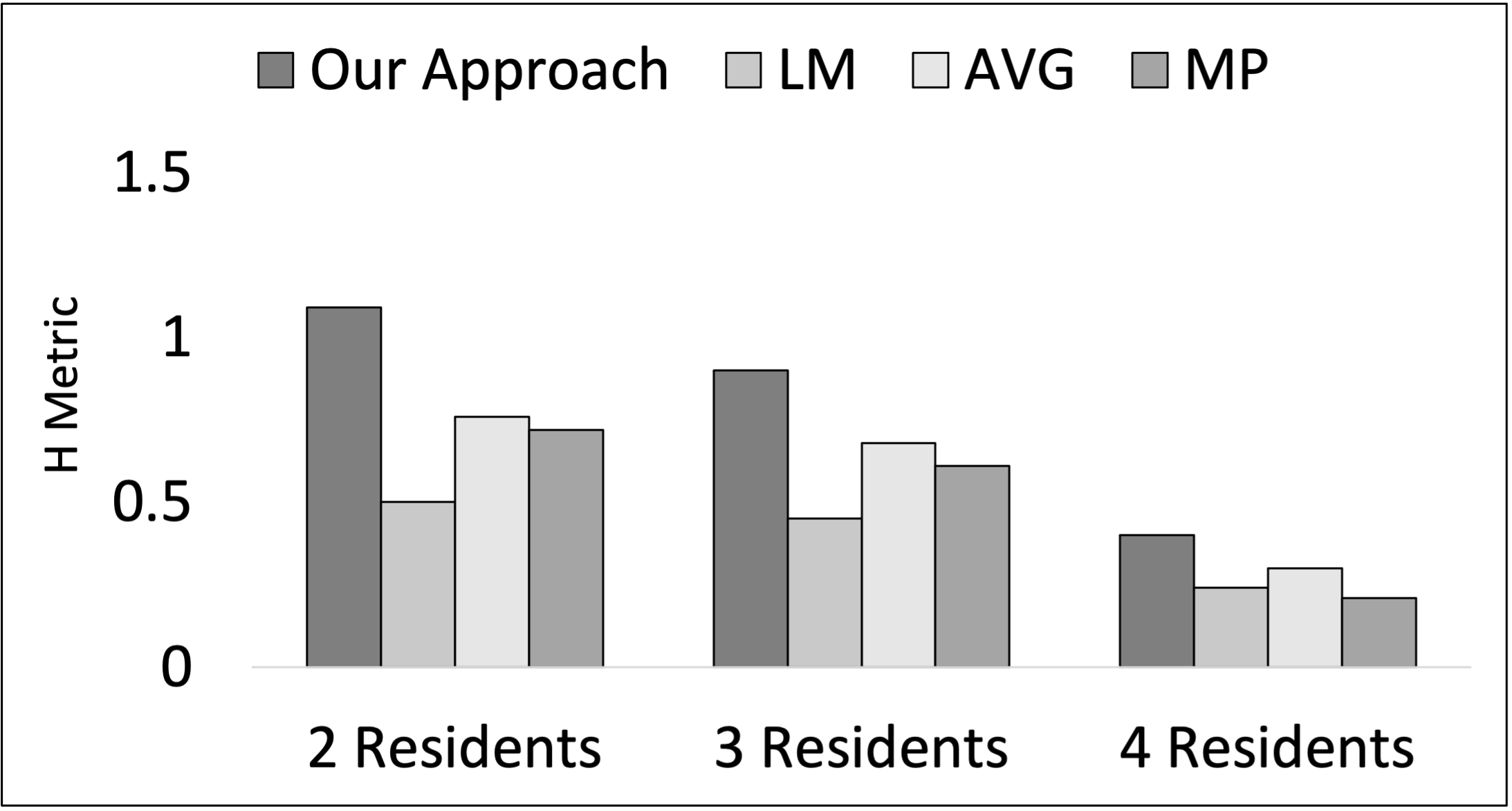} }}%
    \vspace{-2mm}
    \caption{Effectiveness results on CASAS dataset.}%
    %\vspace{2mm}
    \label{casas}%
\end{figure}

\begin{figure}
    \centering
    \subfloat[\centering Satisfaction gain]{{\includegraphics[width=5.5cm,height=3.5cm]{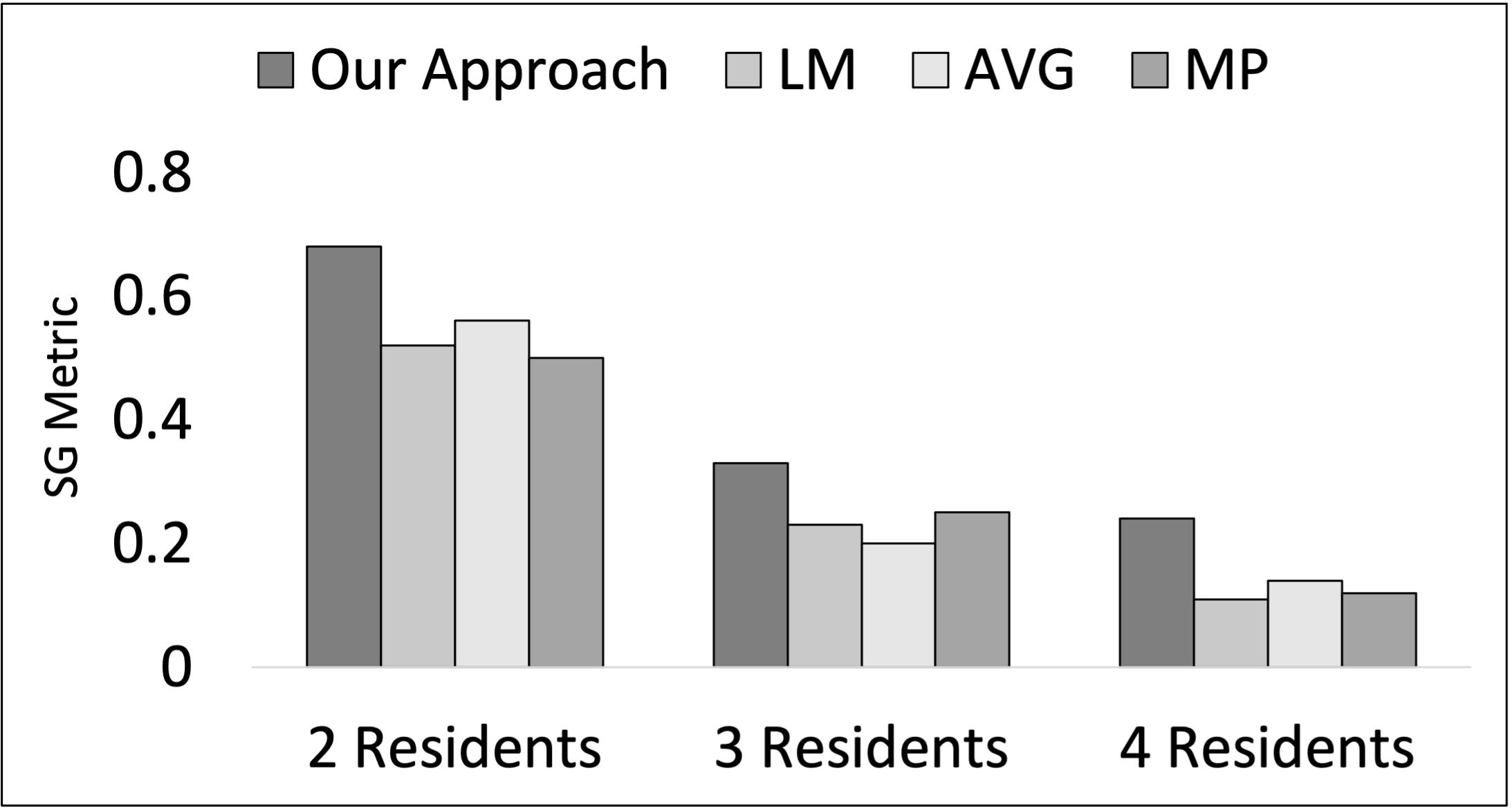} }}
    \qquad
    \subfloat[\centering Harmonic]{{\includegraphics[width=5.5cm,height=3.5cm]{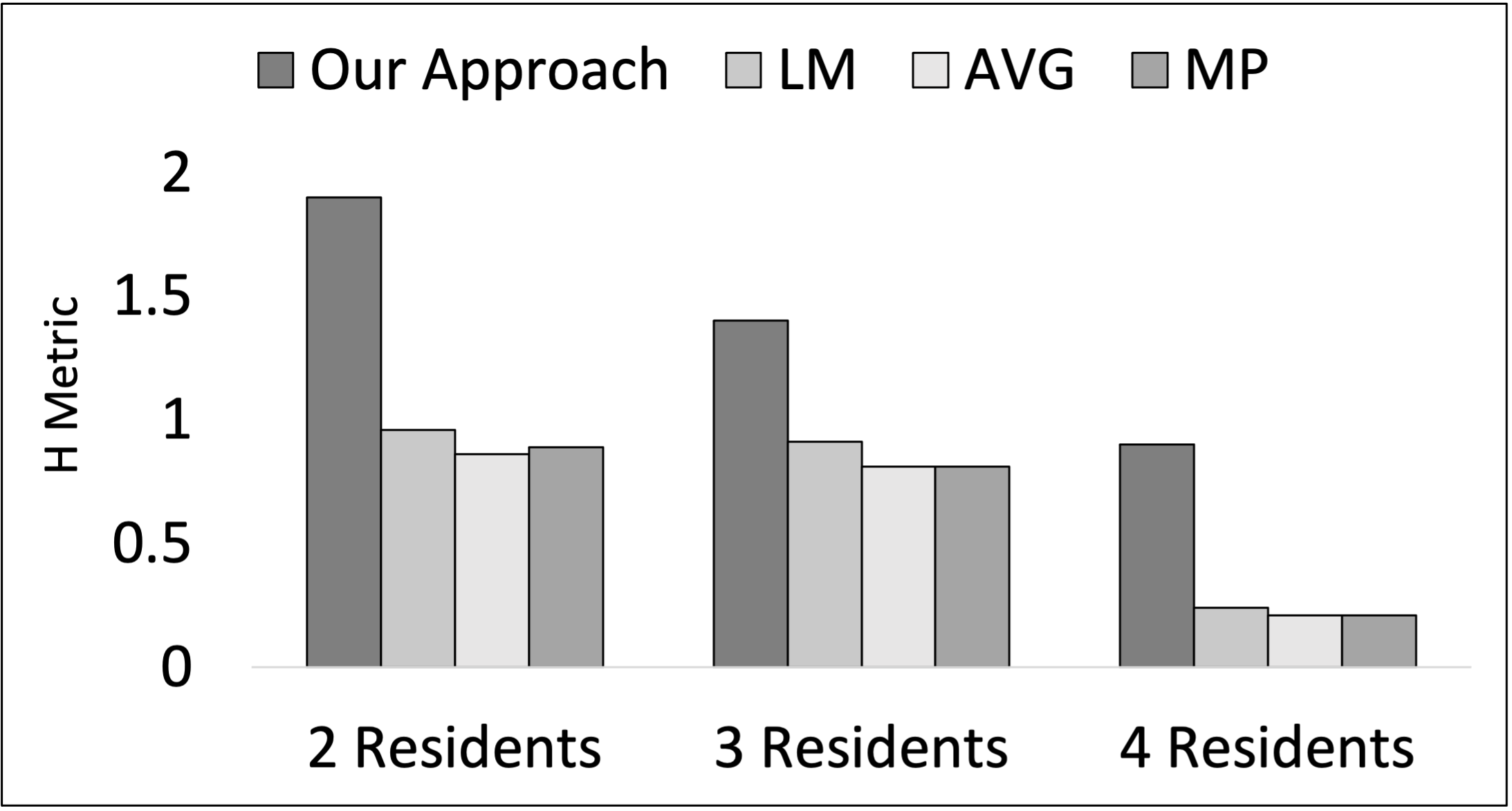} }}
    \vspace{-2mm}
    \caption{Effectiveness results on CAMRa2011 dataset.}
    %\vspace{2mm}
    \label{camra}
\end{figure}

The performances of various conflict resolution strategies are evaluated in this part. The results on two metrics concerning the number of residents are shown in Fig. \ref{casas} and Fig. \ref{camra}.

On SG metric, our approach performs better than other existing approaches with all sizes of groups (Fig. \ref{casas}(a) and Fig. \ref{camra}(a)). AVG does not always perform best because only the adopted items are considered during the computation of SG values. Items with high preference scores are defined as adopted items, meaning those items are frequently used previously. Some items will not be accepted by all the members even though they have high preference scores by other members. In this regard, AVG may lose some gains. Fig. \ref{casas}(b) and Fig. \ref{camra}(b) report the results of different methods based on harmonic metric. Harmonic metric decreases when the group size becomes larger, denoting low fairness in larger groups. Almost all methods perform better on CAMRa2011 dataset than CASAS dataset. However, their performances worsen when the number of residents increases because it is more difficult to aggregate preferences in larger groups.
%The reason is that the groups in CAMRa2011 are much more homogeneous (families) and smaller than the groups in Yelp, and the agreement in homogeneous and small-size groups is more easy to achieve.

%\vspace{5mm}

\begin{figure}
    \centering
    \subfloat[\centering Avg. satisfaction between residents]{{\includegraphics[width=5.5cm,height=3.5cm]{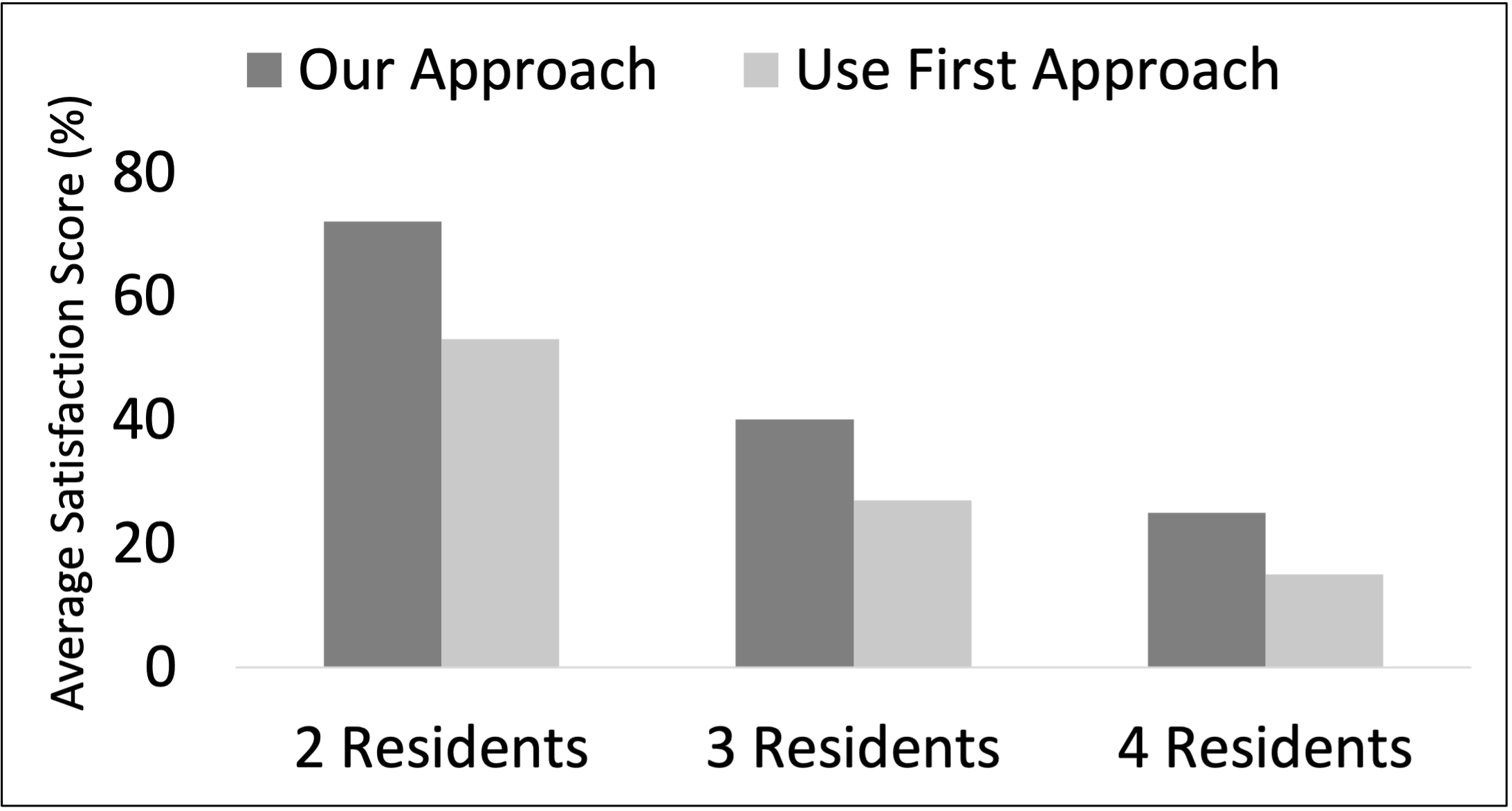} }}
    \qquad
    \subfloat[\centering Avg. satisfaction vs no. of conflicts]{{\includegraphics[width=5.5cm,height=3.5cm]{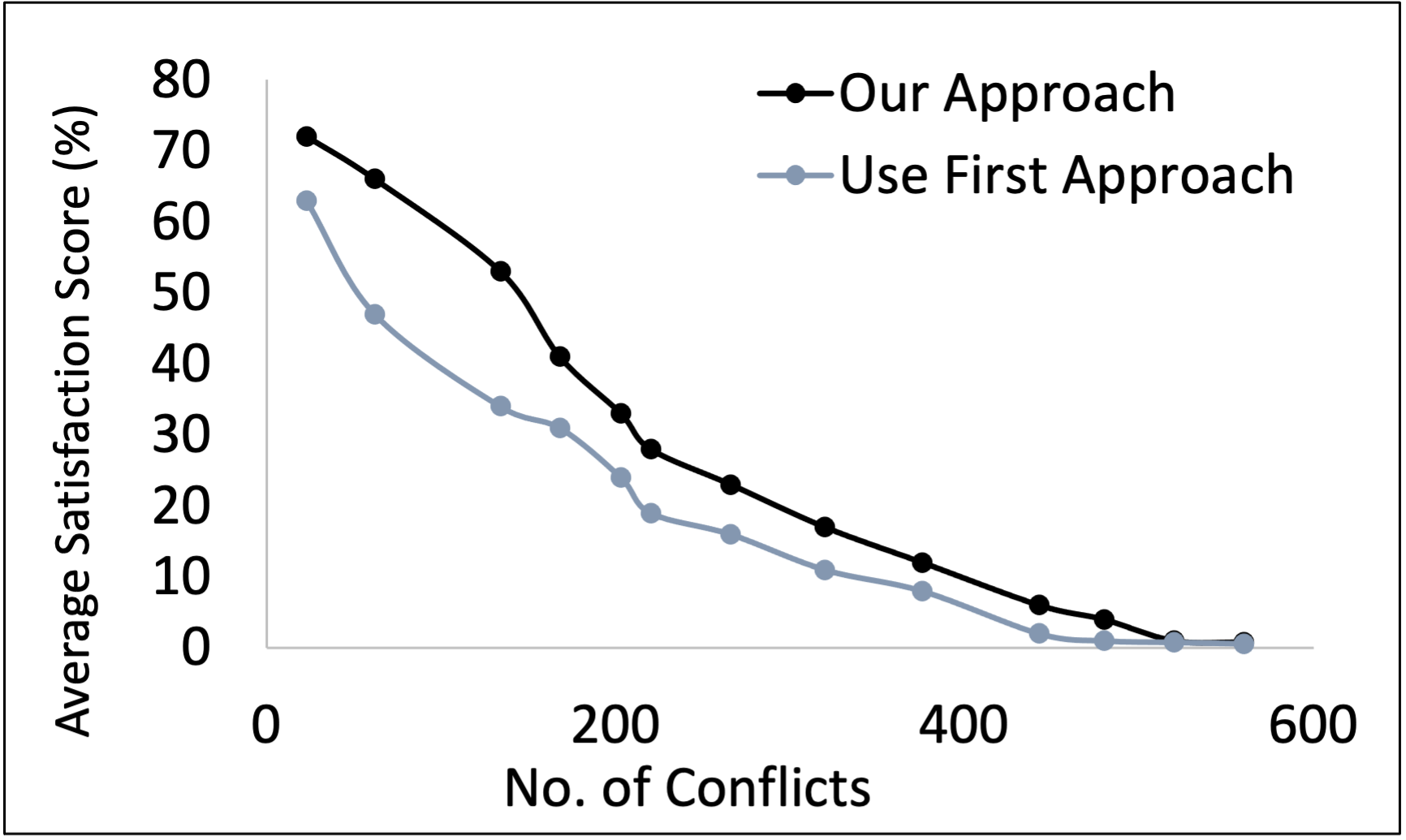} }}
    %\vspace{-1mm}
    \caption{Comparison between our approach and use first approach.}
    %\vspace{2mm}
    \label{avgSatisfaction}
\end{figure}

Finally, we conduct another set of experiments to compare our approach with an existing approach, namely, Use First (UF) proposed in \cite{lee2019situation}. For this experiment, we consider TV channel data to measure satisfaction between residents. We undertake this experiment considering conflicts between 2 residents, 3 residents, and 4 residents, respectively. Fig. \ref{avgSatisfaction} refers that the satisfaction score decreases as the number of residents increases. More residents mean more service requirements, thus creates more service conflicts—the greater number of conflicts, the lesser satisfaction scores. On one hand, in the UF approach, the user who starts watching TV first will be enjoying the TV without considering other residents' preferences. On the other hand, our approach is preemptive. Thus, it resolves conflict by selecting the TV channel that suits most users.

%We use the method described in the 4.3.2 section to calculate satisfaction scores. The CASAS dataset consists of ``watch\_TV" human activities, but it does not specify any channel information, so we augment it.

%\vspace{-4mm}
\section{Related Work}
%\vspace{-3mm}
The concepts of conflict detection and resolution are surveyed in the relevant literature. Conflicts are categorized based on three criteria: (i) source, (ii) intervenience, and (iii) solvability. Different types of sources are responsible for conflict occurrence \cite{ibrhim2021conflicts}. A conflict may occur when many users try to use a resource-defined as a \textit{resource-level conflict} \cite{lalanda2017conflict}. A conflict may happen when several applications utilize a resource simultaneously-regarded as an \textit{application-level conflict} \cite{lalanda2017conflict}. A conflict may arise due to conflicting policies for a given context, known as a \textit{policy-level conflict} \cite{miandashti2020empirical}. Conflicts may arise due to intervenience \cite{shahzaad2020game}. Conflict is common in multi-occupant homes, however, a conflict may happen in single-occupant homes. For instance, a conflict may occur based on contradictory intentions like saving energy and comfort at the same time \cite{lakhdari2020fluid}.

%Solvability is another key criterion for conflict categorization \cite{shahzaad2020game}. Usually conflicts occur in runtime. Sometimes a system cannot detect conflict in runtime and later realizes that it happened because of delayed sensor information.

Some preference aggregation strategies such as average (AVG), least-misery (LM), and most-pleasure (MP) are used for conflict resolution in existing research \cite{cao2018attentive, guo2020group,fattah2018cp}. However, they did not consider the service requirements of the present situation. Only considering current requirements (i.e., interactions), some conflict resolution strategies such as fair principle, use first, and static priority assignment is used in existing research \cite{nurgaliyev2017improved, lee2019situation}. Fairness and interactions are ignored in these works. Consequently, they can not always generate a fair solution for each resident in a conflicting situation, leading to low satisfaction. Hence, there is a need for a framework that ensures fairness by integrating current interactions with preferences extracted from the past usage patterns. We use both previous usage data and current interaction data to build the conflict resolution framework. The proposed framework resolves conflicts aiming to maximize the residents' overall satisfaction.

\vspace{-3mm}
\section{Conclusion and Future Work}
\vspace{-1mm}

We propose a novel approach for conflict resolution of IoT services by combining current interactions and historical interactions. The proposed preference estimation model is developed based on the temporal proximity strategy. The framework employs a preference aggregation model based on singular value decomposition. The effectiveness of the proposed approach is tested with other existing approaches. In our future work, we will improve the conflict resolution framework by utilizing not only preferences, but also other contextual information related to the residents. Factors such as interpersonal relationship can play a vital role for conflict resolution. Meanwhile, we will test our solutions in more complicated scenarios, e.g., more experimental settings on even larger datasets. 
\bibliographystyle{splncs04}
%{\footnotesize
\bibliography{ConfRes}




% LNCS Homepage, \url{http://www.springer.com/lncs}. Last accessed 4
% \end{thebibliography}
\end{document}